\documentclass[twocolumn]{aastex631} 

\usepackage{graphics,epsf}
\usepackage[utf8]{inputenc}
\usepackage{amsmath}                
\usepackage{amsfonts}               
\usepackage{amssymb}                
\usepackage{epsfig}                 
\usepackage{graphicx}               
\usepackage{float}
\usepackage{color}
\usepackage{multirow}               
\usepackage{hyperref}

\hypersetup{
    colorlinks=true,
    linkcolor=red,   
    urlcolor=cyan}

\usepackage[colorinlistoftodos]{todonotes}

\newcommand{\cm}{{~\rm cm}}
\newcommand{\km}{{~\rm km}}
\newcommand{\s}{{~\rm s}}

\newcommand{\g}{{~\rm g}}
\newcommand{\G}{{~\rm G}}
\newcommand{\K}{{~\rm K}}
\newcommand{\erg}{{~\rm erg}}

\begin{document}

\title{Simulating the shaping of point-symmetric structures in the jittering jets explosion mechanism}
\date{March 2025}

\author[0009-0001-4877-1125]{Jessica Braudo}
\affiliation{Department of Physics, Technion - Israel Institute of Technology, Haifa, 3200003, Israel; jessicab@campus.technion.ac.il; amichaelis@campus.technion.ac.il; 
akashi@technion.ac.il;
soker@physics.technion.ac.il}

\author[0000-0002-1361-9115]{Amir Michaelis} 
\affiliation{Department of Physics, Technion - Israel Institute of Technology, Haifa, 3200003, Israel; jessicab@campus.technion.ac.il; amichaelis@campus.technion.ac.il; 
akashi@technion.ac.il;
soker@physics.technion.ac.il}

\author[0000-0001-7233-6871]{Muhammad Akashi} 
\affiliation{ Kinneret College on the Sea of Galilee, Samakh 15132, Israel}
\affiliation{Department of Physics, Technion - Israel Institute of Technology, Haifa, 3200003, Israel; jessicab@campus.technion.ac.il; amichaelis@campus.technion.ac.il; 
akashi@technion.ac.il;
soker@physics.technion.ac.il}

\author[0000-0003-0375-8987]{Noam Soker}
\affiliation{Department of Physics, Technion - Israel Institute of Technology, Haifa, 3200003, Israel; jessicab@campus.technion.ac.il; amichaelis@campus.technion.ac.il; 
akashi@technion.ac.il;
soker@physics.technion.ac.il}

\begin{abstract}
We conduct three-dimensional hydrodynamical simulations of core-collapse supernovae by launching several pairs of jets into a collapsing core model and show that the jittering jets explosion mechanism (JJEM) can form a point-symmetric morphology that accounts for observed morphologies of about a dozen core-collapse supernovae (CCSN) remnants. Point-symmetric morphologies are composed of pairs of opposite structures around the center of the CCSN remnant. In the JJEM, the newly born neutron star launches several to a few tens of pairs of jets with stochastically varying directions, and these jets explode the star. In the simulations with the FLASH numerical code, we launch pairs of jets with varying directions, energies, opening angles, and durations into the massive stellar core and follow their evolution for about two seconds. We show that the jets form pairs of opposite filaments, clumps, bubbles, and lobes, namely, prominent point-symmetric morphologies. The interaction of the jets with the core leads to vigorous Rayleigh-Taylor instabilities and excites many vortices, which also shape clumps and filaments. Our results suggest that the JJEM could play a central role in the explosion mechanism of CCSNe; neutrino heating can boost the role of jets.
\end{abstract}

\keywords{supernovae: general -- stars: jets -- stars: massive -- ISM: supernova remnants}


\section{Introduction}
\label{sec:Introduction}

The gravitational energy of the collapsing core powers core-collapse supernovae (CCSNe). Neutrinos carry most of the released gravitational energy as the neutron star (NS) cools down. A small fraction of this energy explodes the star. Two alternative theoretical models consider two different processes to deliver this energy to explode the star (for recent reviews, see \citealt{Soker2024UnivReview, Janka2025}). 

In the delayed neutrino explosion mechanism, a small fraction of the neutrinos interact with the collapsing core in the gain region, at $\approx 100 \km$ from the newly born NS, heat the material, revive the stalled shock and explode the star (e.g., \citealt{Andresenetal2024, BoccioliFragione2024, Burrowsetal2024kick, JankaKresse2024, Muler2024, Mulleretal2024, vanBaaletal2024, WangBurrows2024, Laplaceetal2024, Huangetal2024, Bocciolietal2025, Bocciolietal2025a,  Imashevaetal2025, Nakamuraetal2025, Janka2025, WangTBurrows}, for some very recent studies).  In the magnetorotational explosion mechanism, in rare cases of CCSNe, those that have a rapidly rotating pre-collapse core (e.g., \citealt{Shibagakietal2024, ZhaMullerPowell2024, Shibataetal2025}, for recent studies of this mechanism), jets along a fixed axis explode the star; however, according to this mechanism, in most CCSNe, neutrino heating explodes the star. Therefore, we consider the magnetorotational explosion mechanism part of the neutrino-driven mechanism, even though, for these rare cases, fixed-axis jets deliver most of the gravitational energy to the exploding star. 
   
In the jittering jets explosion mechanism (JJEM), jets deliver the gravitational energy of the collapsing core to the star and explode it (for a recent detail on the JJEM and its parameters, see  \citealt{Soker2025Learning}). The newly born NS accretes material via intermittent accretion disks; the disks launch pairs of opposite jets with varying directions, i.e., jittering jets.  Only in rare cases, when the pre-collapse core is rapidly rotating, do the pairs of jets have a common axis. The stochastic behavior of the accretion disks results from angular momentum fluctuations in the pre-collapse core's convection zones; instabilities around the newly born NS amplify these fluctuations (e.g., \citealt{GilkisSoker2014, GilkisSoker2016, ShishkinSoker2021, ShishkinSoker2023, WangShishkinSoker2024}). Neutrino heating plays a secondary role in boosting the effect of the jets \citep{Soker2022nu}. Jets supply most of the explosion energy even when the newly born NS is highly magnetized and rapidly rotating, i.e., an energetic magnetar (e.g., \citealt{SokerGilkis2017, Kumar2025}).

One robust outcome of the JJEM is that many, but not all, CCSN remnants should possess point-symmetric morphology. Namely, the CCSN remnant has pairs of structural features opposite to the center of the CCSN remnant, and the pairs do not share their symmetry axes. The opposite structural features might be filaments, clumps, bubbles (faint zones closed by a brighter rim), lobes (bubbles with partial rims), and ears (protrusions from the main remnant's shell having a decreasing cross-section with distance from the center). 
The neutrino-driven explosion mechanism cannot account for the point-symmetric morphologies of about a dozen CCSN remnants. 

The JJEM accounts for the point-symmetric morphology by pairs of jets that shape the supernova ejecta, hence the remnant. 
Many jets are choked inside the core, leaving no imprints on the remnants. Only one to several of the $N_{\rm 2j} \simeq 5-30$ pairs of exploding jets \citep{Soker2025Learning} leave observable marks.
In many cases, the point-symmetric morphology is not sharp and clear because several processes smear it (e.g., \citealt{SokerShishkin2025Vela}), including interaction with circumstellar material, interaction with the interstellar medium, instabilities in the explosion process, the NS natal kick, heating of the ejecta by radioactive decay and the reverse shock, and the pulsar wind nebula. In recent years, studies have identified point-symmetric morphologies in over ten CCSN remnants and connected them to jet shaping, solidifying the JJEM as the primary and even sole explosion mechanism of CCSNe. Some of these point-symmetric CCSN remnants have a pair or more of opposite clumps, e.g.,  
SNR 0540-69.3 \citep{Soker2022SNR0540},
Vela (\citealt{Soker2023SNRclass, SokerShishkin2025Vela}), 
SN 1987A \citep{Soker2024NA1987A, Soker2024Keyhole},
Cassiopeia A \citep{BearSoker2025CasA}, W49B, \citep{SokerShishkin2025W49}, 
and possibly also the Crab Nebula \citep{ShishkinSoker2025Crab}.
Other CCSN remnants do not show prominent pairs of clumps; these are 
N63A \citep{Soker2024CounterJet}, 
G321.3–3.9 \citep{Soker2024CF, ShishkinSoker2024}, 
G107.7-5.1 \citep{Soker2024CF}, 
the Cygnus Loop \citep{ShishkinKayeSoker2024}, 
W44 \citep{Soker2025W44}, 
CTB~1 \citep{BearSoker2023RNAAS}, 
Puppis A \citep{Bearetal2025Puppis}, and SNR~G0.9+0.1 \citep{Soker2025G09}.
In attributing the point-symmetric pairs to shaping by jets, these studies applied a very common tool in the study of planetary nebulae, where studies consider jets to be behind the shaping of multipolar planetary nebulae and other point-symmetric morphologies (e.g.,  \citealt{Morris1987, Soker1990AJ, SahaiTrauger1998, AkashiSoker2018,   EstrellaTrujilloetal2019, Tafoyaetal2019, Balicketal2020, RechyGarciaetal2020, Clairmontetal2022, Danehkar2022, MoragaBaezetal2023, Ablimit2024, Derlopaetal2024, Mirandaetal2024, Sahaietal2024}). The tool of identifying point-symmetric morphologies and connecting them to jet-driven explosions is new in the study of CCSNe. Our study is on that topic.  
 In many planetary nebulae, the jets precess, namely, continuously varying their direction (e.g., \citealt{Guerreroetal1998, Mirandaetal1998, Sahaietal2005, Boffinetal2012, Sowickaetal2017, RechyGarciaetal2019, Guerreoetal2021, Clairmontetal2022}). We postpone the study of precessing jets in CCSNe to a future study. 

Studies in the frame of the neutrino-driven mechanism argue for the successful shaping of different structures in CCSN remnants. For example, several papers address the shaping of Cassiopeia A, mainly the inner parts of the remnant (e.g., \citealt{Orlandoetal2021, Orlandoetal2022, Orlandoetal2025, DeLoozeetal2024}), but these papers do not address the point-symmetric morphology.  

In this study, we simulate the interaction of jittering jets with the core of a stellar model and examine the formation of point-symmetric morphologies just as the jets exit the core. In a follow-up study, we will follow the clumps and the other point-symmetric structures until they break out from the star. \citet{PapishSoker2014Planar} conducted the first and only simulation of this kind, demonstrating the very early formation of clumps by jittering jets. They did not follow the clumps into larger distances or explore the physics involved; these are the goals of the present study. 
There are other simulations of non-constant axis jets propagating in the core of exploding massive stars and aiming at rare explosions that leave a black hole (e.g., wobbling jets: \citealt{Gottliebetal2022, Gottliebetal2023,BoppGottlieb2025}). We differ in having jittering and non-relativistic jets as a very common explosion process.  
We describe our numerical setup in Section \ref{sec:Numerical}, and the results in Sections \ref{sec:Density} and \ref{sec:RTI}. We summarize this first paper in a series in Section \ref{sec:Summary}.

\section{Numerical setup}
\label{sec:Numerical}

\subsection{Grid and initial conditions}
\label{subsec:Grid}


We conduct our three-dimensional (3D) hydrodynamical simulations using version 4.8 of the {\sc flash} code \citep{Fryxell2000}, employing the unsplit hydro solver \citep{2006PhDT.......113L,2009JCoPh.228..952L,2013JCoPh.243..269L}. FLASH is a modular, adaptive-mesh refinement (AMR) code designed for solving hydrodynamics and magnetohydrodynamics problems.  

We utilize a fully three-dimensional AMR grid with 8 refinement levels, resulting in $2^{10}$ cells along each axis. The computational domain is a Cartesian grid $(x,y,z)$ with outflow boundary conditions applied at all boundaries. The plane $z=0$ is set as the equatorial plane of both the star and the surrounding medium. Our simulations cover the entire spatial domain, including both sides of the equatorial plane.  

The grid extends over a cubic region of $(60,000\km)^{3}$ in our simulations. We use the same density profile of the collapsing core as \citet{PapishSoker2014Planar} adapted from \citet{Liebendorferetal2005} for a $15 M_\odot$ star at $t\simeq 0.2\s$ after bounce. The density profile spans over radius values of $r\simeq6-30000\km$ and density values in the range $\rho \simeq 3\times10^3 - 4\times 10^{14} \ \rm{g \ cm^{-3}}$.

We include the gravity of a point mass of $1.4 M_\odot$ at the center to represent the newly born NS. This gravitational potential influences the surrounding gas, particularly affecting the backflow. For numerical reasons that allow us to conduct six simulations, we inject the jets at $500 \km$ and not at $50 \km$. We also do not include NS wind or later jets that result from the accretion of the back-flowing material. For these reasons, in this first study in a series, we do not analyze the flow near the center, namely the inner $\simeq 15\%$ of the grid size, which corresponds to the volume inner to $\simeq 5000 \km$. 
According to the JJEM, the accretion disk (or belt) launches the jets at $20-50 \km$ from the center. The jets interact with the inflowing core material both below and above the stalled shock, in the region of $\simeq 50-200 \km$. This turbulent inflow, along with the unrelaxed intermittent accretion disk that launches the jets, leads to the two opposite jets of each jet-launching episode not being equal when they emerge at hundreds of kilometers from the center. Other than that, there are no large effects on the role of the jets in exploding the core. Therefore, injecting the jets at $500 \km$ is adequate for this study. In this study, we inject equal jets in each pair. In future simulations, we will inject two unequal opposite jets to better reflect the expected behavior according to the JJEM.


We use the electron-positron Helmholtz free energy Equation of State (EOS) \citep{2000ApJS..126..501T, 1999ApJS..125..277T, 1998ApJS..117..627A}. Assuming temperature and density as the free variables and employing the Fermi-Dirac distribution for a noninteracting Fermi gas of ionized matter, we can determine the number density and all derived thermodynamic quantities. At low densities and temperatures, this EOS coincides with the gamma equation of state while also accounting for denser matter where degeneracy plays a significant role. Additionally, this implementation incorporates Coulomb corrections and radiation blackbody pressure, making it more suitable for this simulation.

\subsection{The jittering jet pairs}
\label{subsec:Jets}

The jets are injected into the computational domain using a predefined injection scheme based on spherical coordinates $(r, \theta, \phi)$. Each jet is introduced in a cone within an injection radius $r_{\rm j}=500 \km$ and a half-opening angle $\alpha_{\rm j}$. The direction of the jet launching is determined by the angles $\theta_{\rm j}$ and $\phi_{\rm j}$ in the spherical coordinate system, where $\theta_{\rm j}$ is the polar angle measured from the $z$-axis, and $\phi_{\rm j}$ is the azimuthal angle in the $(x,y)$ plane measured from the $x$-axis.

The jets' launching velocity is $v_{\rm j} = 5 \times 10^{4} \km \s^{-1}$, approximately the escape velocity of the still relaxing newly born neutron star. Their initial thermal energy is $\simeq 20\%$ of their initial kinetic energy (following \citealt{PapishSoker2014Planar}). 

\subsection{Simulated cases}
\label{subsec:Cases}

We simulate six cases. Section \ref{subsec:Jets} describes the common properties of the jets. We give the parameters that we vary between the simulations in six tables (Table \ref{Tab:Table3} - \ref{Tab:Table6} are in \ref{subsec:initial_param}). We focus on A3 and A5. 

Simulation A3 consists of three pairs of jets launched sequentially, forming a pyramid shape. Table \ref{Tab:Table1} lists the parameters of the pairs. The jets have the same half-opening angle $\alpha_{\rm j}$, velocity $v_{\rm j}$, and mass loss rate $\dot m_{\rm j}$. Note that the last pair lasts longer than the other pairs. The total kinetic energy carried by the six jets in this simulation is $E_{\rm k, tot} = 2.5 \times 10^{51}\ \rm erg$ (the thermal energy adds $20\%$ of the kinetic energy). 
Simulation A5 consists of five pairs of jets, with their properties listed in Table \ref{Tab:Table2}. The first three pairs in this simulation are identical to those in simulation A3. There is a quiet period between consecutive jet-launching episodes. Note that the last pair lasts longer and has a larger half-opening angle than the other pairs. The total kinetic energy carried by the ten jets in this simulation is $E_{\rm k, tot} = 1.5 \times 10^{51}\ \rm erg$. 
\begin{table}[h]
\scriptsize
\begin{center}
  \caption{Initial parameters of simulation A3}
    \begin{tabular}{| p{1.8cm} | p{1.1cm} | p{1.1cm}| p{1.1cm} |}

\hline
Jets' pair & $1$ & $2$ & $3$ \\
\hline
$t_{\rm j}({\rm start})\ [\rm s]$ & $0.00$ & $0.05$ & $0.1$\\
\hline
$t_{\rm j} {\rm(end)}\ [\rm s]$ & $0.05$ & $0.1$ & $0.25$\\
\hline
$\theta_{\rm j}\ [\rm deg]$ & $50$ & $50$ & $50$ \\
\hline
$\phi_{\rm j}\ [\rm deg]$ & $0$ & $120$ & $240$ \\
\hline
$\alpha_{\rm j}\ [\rm deg]$ & $10$ & $10$ & $10$ \\
\hline
$\dot{m}_{\rm j}\ [10^{32} \g \s^{-1}]$ & $4$ & $4$ & $4$ \\
\hline
$E_{\rm k, 2j}\ [10^{50} \erg]$ & $5$ & $5$ & $15$ \\
\hline

\end{tabular}
  \\
\label{Tab:Table1}
\end{center}
\begin{flushleft}
\small 
Notes: The initial parameters of simulation A3, which has three pairs of jets. In all simulations, the launching velocity of the jets is $v_{\rm j} = 50,000 \km \s^{-1}$. The second and third rows provide the initial and final times for each jet-launching episode.  $(\theta_{\rm j}, \phi_{\rm j})$ is the direction of the symmetry axis of the pair, where $\theta_{\rm j}$ is the polar angle, i.e., the angle to the $z$ axis in our grid, and $\phi_{\rm j}$ is the azimuthal angle, i.e., the angle of the projected axis on the $xy$ plane relative to the $x$ axis;  $\alpha_{\rm j}$ is the half opening angle of the jets; $\dot{m}_{\rm j}$ is the mass loss rate of one jet; $E_{\rm k,2j}$ is the kinetic energy of the pair of jets; thermal energy is $20\%$ of that. Note that the last pair lasts longer than the other pairs. The total kinetic energy carried by the jets in this simulation is $E_{\rm k, tot} ({\rm A3}) = 2.5 \times 10^{51}\ \rm erg$.   
\end{flushleft}
\end{table}
\begin{table}[h]
\scriptsize
\begin{center}
  \caption{Initial parameters of simulation A5}
    \begin{tabular}{| p{1.8cm} | p{0.7cm} | p{0.7cm}| p{0.7cm}| p{0.7cm} | p{0.7cm} |}

\hline
Jets' pair& $1$ & $2$ & $3$& $4$ & $5$ \\
\hline
$t_{\rm j}({\rm start})\ [\rm s]$ & $0.00$ & $0.07$ & $0.14$ & $0.21$ & $0.28$ \\
\hline
$t_{\rm j} {\rm(end)}\ [\rm s]$ & $0.05$ & $0.12$ & $0.19$ & $0.26$ & $0.413$ \\
\hline
$\theta_{\rm j}\ [\rm deg]$ & $50$ & $50$ & $50$ & $0$ & $0$ \\
\hline
$\phi_{\rm j}\ [\rm deg]$ & $0$ & $120$ & $240$ & $0$ & $20$\\
\hline
$\alpha_{\rm j}\ [\rm deg]$ & $10$ & $10$ & $10$ & $10$ & $20$ \\
\hline
$\dot{m}_{\rm j}\ [10^{32} \rm g \s^{-1}]$ & $1.8$ & $1.8$ & $1.8$ & $1.8$ & $1.8$ \\
\hline
$E_{\rm k,2j}\ [10^{50} \ \rm erg]$ & $2.25$ & $2.25$ & $2.25$ & $2.25$ & $6$ \\
\hline

\end{tabular}
  \\
\label{Tab:Table2}
\end{center}
\begin{flushleft}
\small 
Notes: Similar to Table \ref{Tab:Table1} but for simulation A5, composed of five pairs of jets with a total jets' kinetic energy of $E_{\rm k, tot} ({\rm A5}) = 1.5 \times 10^{51}\ \rm erg$.    
Note that there are quiet periods between pairs; the last pair has a larger half-opening angle $\alpha_{\rm j}$ and lasts longer than the other pairs. 
\end{flushleft}
\end{table}

The other four simulations differ in several parameters from simulations A3 and A5.
Since we present fewer graphs of the results for these simulations, we present their parameters in Tables \ref{Tab:Table3} - {\ref{Tab:Table6} in Appendix \ref{subsec:initial_param}.

Simulations P1 and P2 consist of three pairs of jets launched sequentially in the $y=0$ plane. The first pair in simulation P2 has a larger half-opening angle $\alpha_{\rm j}$ than in P1. The last pair in both simulations lasts longer than the other pairs. The total kinetic energy carried by the six jets in each of these simulations is $E_{\rm k, tot} = 2.5 \times 10^{51}\ \erg$. 

Simulation B5 consists of five pairs of jets and is identical to A5 except that the last pair's half-opening angle $\alpha_{\rm j}$ is smaller. The total kinetic energy the ten jets in this simulation carry is $E_{\rm k, tot} = 1.5 \times 10^{51}\ \rm erg$.
Simulation B7 consists of seven pairs of jets. The first five pairs are identical to those in simulation A5. All pairs have the same active time of jet-launching. The last pair has a smaller mass loss rate $\dot{m_{\rm j}}$. The total kinetic energy the jets carry in this simulation is $E_{\rm k, tot} = 1.5 \times 10^{51}\ \rm erg$.

\section{Results: The point-symmetric density structure}
\label{sec:Density}
\subsection{Density structures}
\label{subsec:Density}

In analyzing the results, we will not refer to the center of the grid, around the jets' launching zone, a radius of about $15\%$ of the size of the grid (inner $\simeq 5000 \km$), because we have not yet let the core fully expand and have not considered the backflow near the center, which might lead to later jets. We will follow the explosion to much later times and consider possible later jets by the backflow that the jets induce in a future paper in the series.  

Figure \ref{Fig:densA3} presents the density of simulation A3 (see Table \ref{Tab:Table1}) in three planes, from top to bottom $z=0$, $x=0$, and $y=0$, and at three times as indicated in the upper row. Figure \ref{Fig:densA5} presents the density maps of simulation A5; note the different density scales of the color bars and the different times of the maps. The total energy in simulation A3 is larger than in simulation A5, hence it evolves faster.   
\begin{figure*}
\begin{center}
\vspace{-0.0cm}
\includegraphics[trim=1.0cm 5.0cm 0.0cm 5.0cm ,clip, scale=0.95]{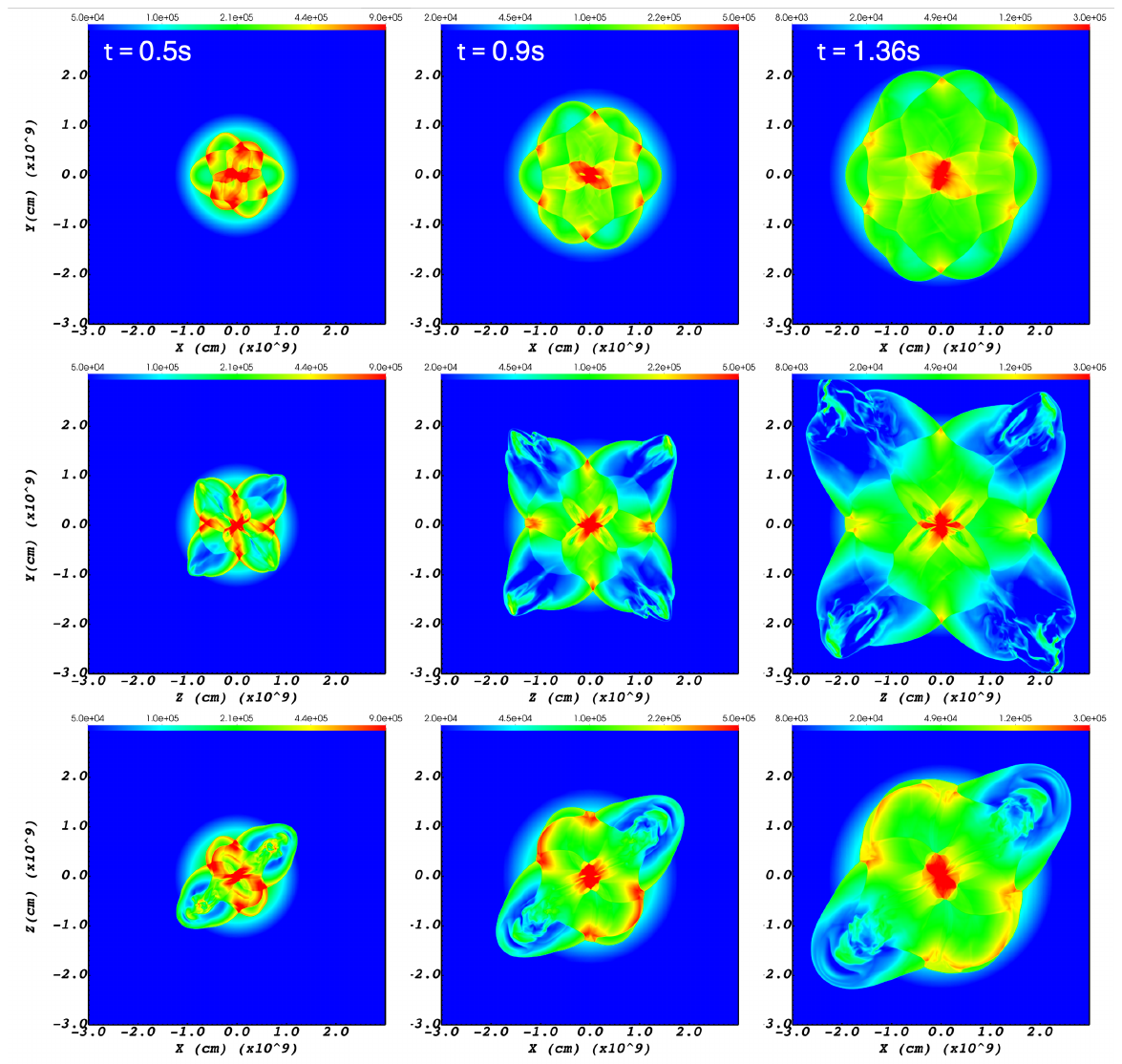} 
\vspace{-0.2cm}
\caption{Density color maps of simulation A3 at three planes, from top to bottom, $z=0$, $x=0$ and $y=0$, at three times.
The density scales on the upper color bar differ between columns. 
Left column: $t=0.5\s$, $0.25 \s$ after the last pair of jets was launched; density scale from deep blue to deep red is $5 \times 10^4 - 9 \times 10^5 \g \cm^{-3}$. Middle panels: $t=0.9\s$; $2 \times 10^4 - 5 \times 10^5 \g \cm^{-3}$. Right panels: $t=1.36\s$, the end of the simulation; $8 \times 10^3 - 3 \times 10^5 \g \cm^{-3}$. 
}
\label{Fig:densA3}
\end{center}
\end{figure*}
\begin{figure*}
\begin{center}  
\vspace{-0.0cm}
\includegraphics[trim=1.0cm 5.0cm 0.0cm 5.0cm ,clip, scale=0.95]{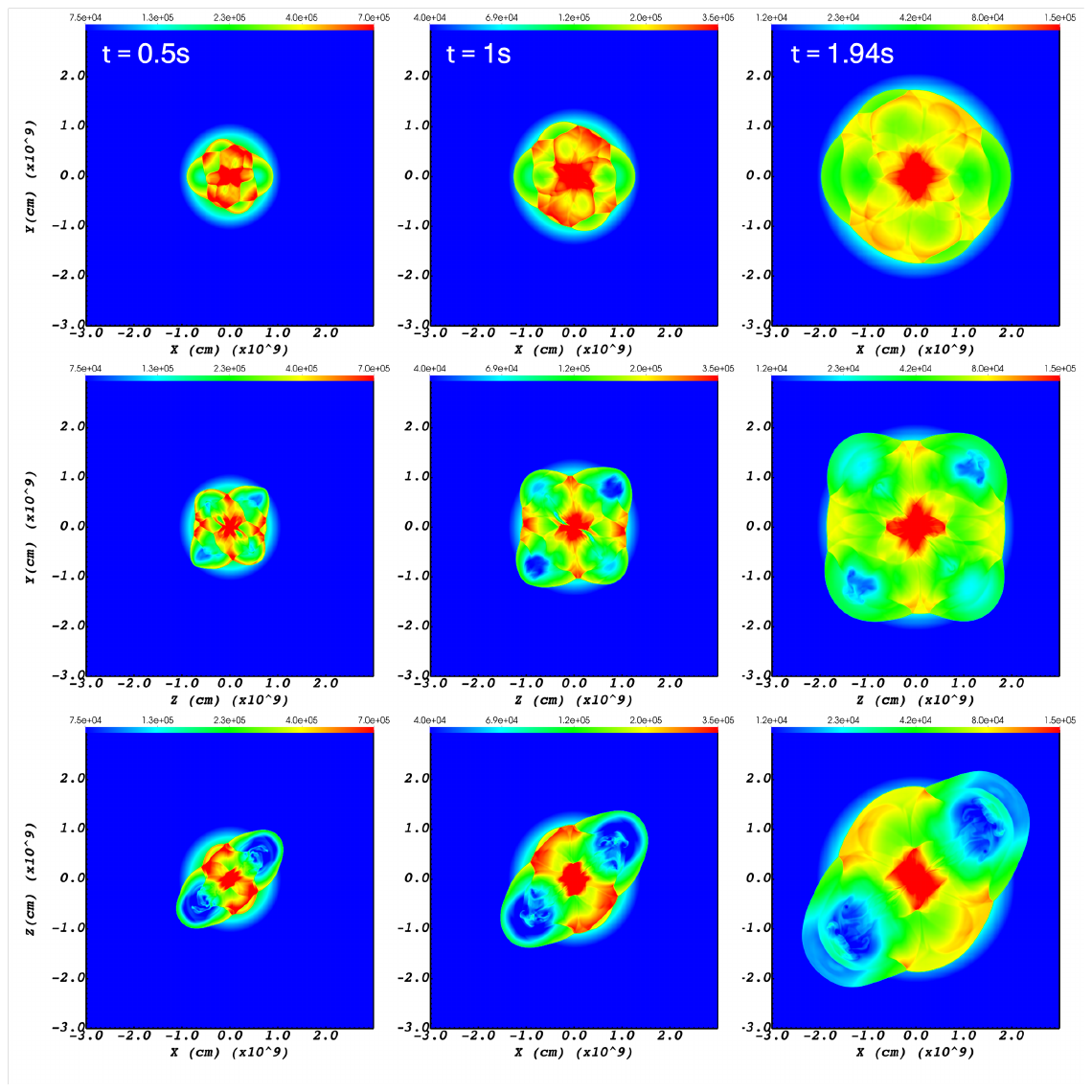} 
\vspace{-0.2cm}
\caption{Similar to figure \ref{Fig:densA3} but for simulation A5 and at different times. Left column: $t=0.5\s$, immediately after the last pair of jets was launched; density scale from deep blue to deep red is $7.5 \times 10^4 - 7 \times 10^5 \g \cm^{-3}$. Middle panels: $t=1\s$; $4 \times 10^4 - 3.5 \times 10^5 \g \cm^{-3}$. Right panels: $t=1.94\s$, the end of the simulation; $1.2 \times 10^4 - 1.5 \times 10^5 \g \cm^{-3}$. 
}
\label{Fig:densA5}
\end{center}
\end{figure*}

Figures \ref{Fig:densA3} and \ref{Fig:densA5} present the point-symmetric density structure that the jets shape in the core. \citet{PapishSoker2014Planar} already obtained similar results, but we use a larger grid and more cases. The 2D density maps in the planes show bubbles (low-density zones closed with a high-density rim), high-density elongated structures, i.e., filaments, and high-density clumps, in a point-symmetric morphology. Namely, each structural feature has a partner on the other side of the center. 
The last jet activity in simulation A3 was at $t_{\rm f,j}=0.25 \s$ and in simulation A5 at $t_{\rm f,j}=0.413 \s$. Figures \ref{Fig:densA3} and \ref{Fig:densA5} show that the point-symmetric morphologies in the two simulations remain prominent for five times as long. The point-symmetric structure reaches distances from the center of $\simeq 20,000-30,000 \km$. Beyond those distances, the density sharply drops and we expect the point-symmetric structures to stay in the CCSN remnant phase. Many processes will smear the point-symmetric structure, including the NS kick, instabilities, interaction with the circumstellar material, and heat release by radioactive decay that forms bubbles (nickel bubbles). Therefore, identifying point-symmetric morphologies in CCSN remnants might be challenging (see references in Section \ref{sec:Introduction}).  

To further illustrate the complex point-symmetric morphologies, in Figures \ref{Fig:contouryx} - \ref{Fig:contourzx}  we present 3D maps of the six simulations from three different points of view. Each 3D map presents three equidensity surfaces, in yellow (low density), red, and blue (high density; see captions). The colored surfaces are partially transparent, so one can see deeper surfaces. Because the colored-surfaces are not fully transparent, zones away from the observer appear differently than those closer to the observer. Therefore, in some cases, there is no full symmetry between opposite sides in those maps. 
\begin{figure*}[t]
\begin{center}
\vspace{-0.0cm}
\includegraphics[trim=0.0cm 0.0cm 0.0cm 0.0cm ,clip, angle=0, scale=0.6]{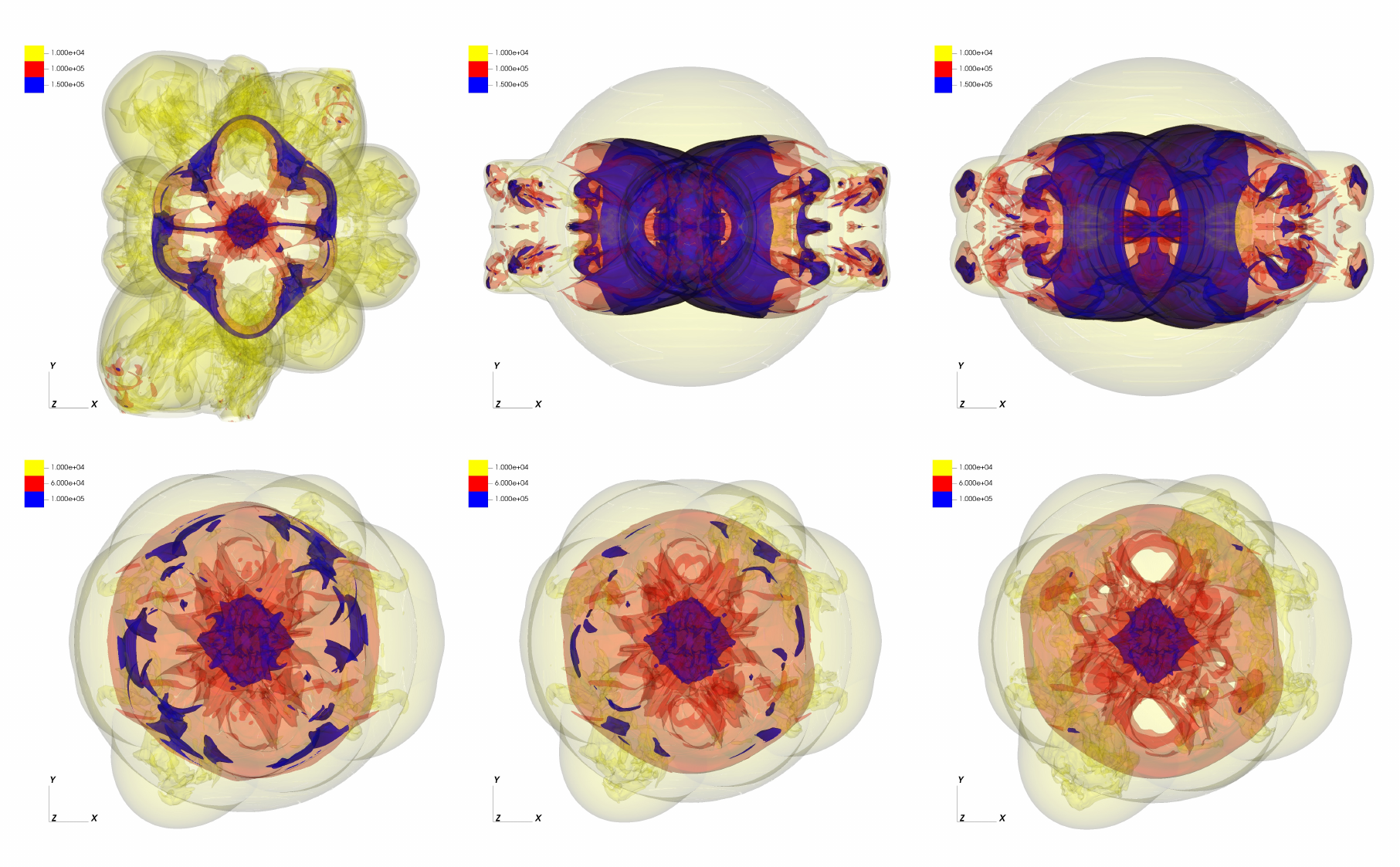} 
\vspace{-0.0cm}
\caption{3D density maps of the six simulations seen from the $z$ direction. Each map presents three equidensity surfaces, partially transparent. The density surfaces in the upper row are $\rho = 10^{4}\g \cm^{-3}$ (yellow), $\rho = 10^{5}\g \cm^{-3}$ (red), and $\rho = 1.5\times10^{5}\g \cm^{-3}$ (blue), and in the lower row are $\rho = 10^{4}\g \cm^{-3}$ (yellow), $\rho = 6\times10^{4}\g \cm^{-3}$ (red), and $\rho = 10^{5}\g\ \cm^{-3}$ (blue).  
Upper left: A3 at $t=1.36\s$. Upper middle: P1 at $t=0.91\s$. Upper right: P2 at $t=0.91\s$. Lower left: A5 at $t=1.94\s$. Lower middle: B5 at $t=2\s$. Lower right: B7 at $t=2.1\s$. 
}
\label{Fig:contouryx}
\end{center}
\end{figure*}
\begin{figure*}[t]
\begin{center}
\vspace{-0.0cm}
\includegraphics[trim=0.0cm 0.0cm 0.0cm 0.0cm ,clip, angle=0, scale=0.6]{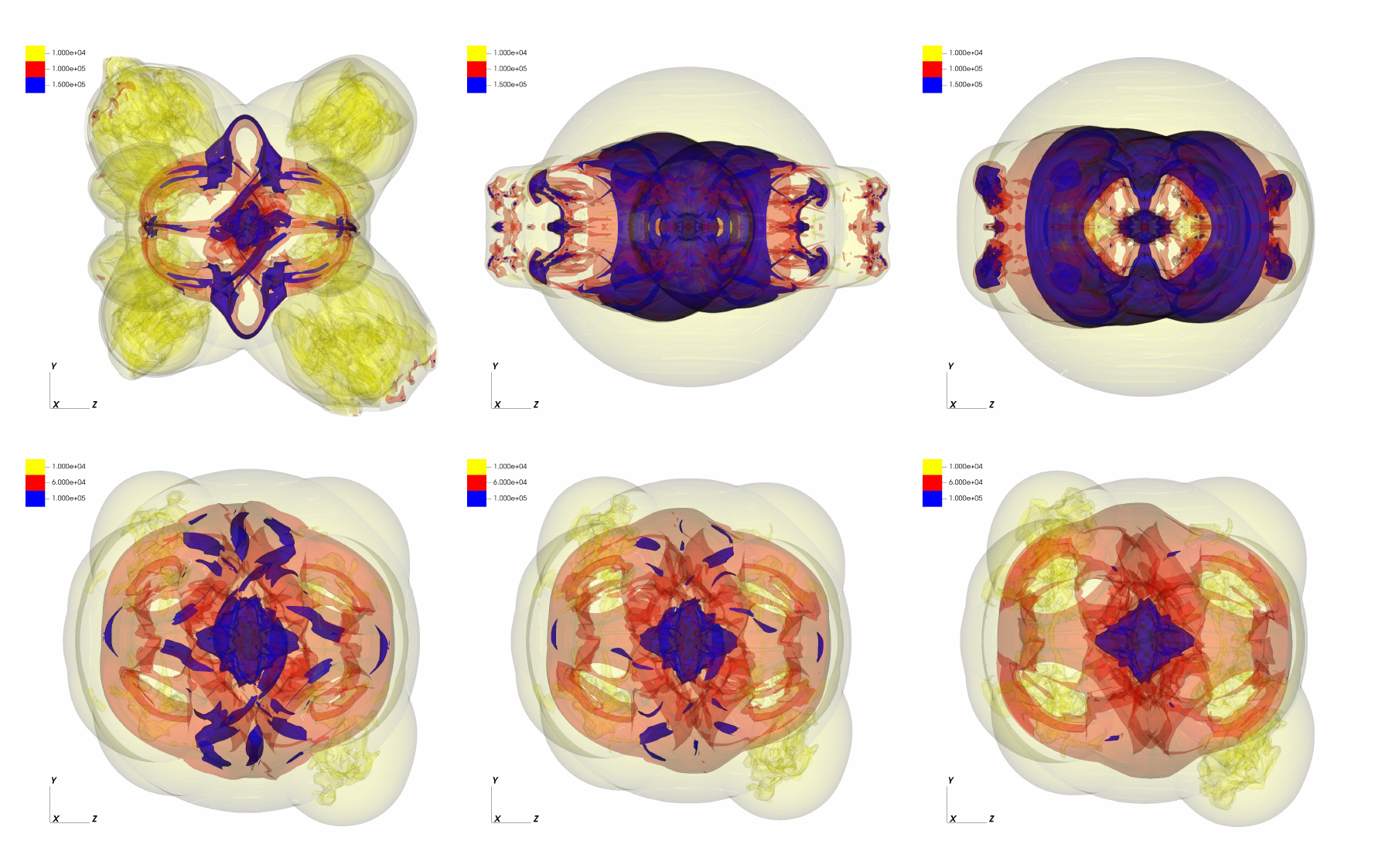} 
\vspace{-0.0cm}
\caption{Similar to Figure \ref{Fig:contouryx} but as seen from the $x$ direction. 
}
\label{Fig:contouryz}
\end{center}
\end{figure*}
\begin{figure*}[t]
\begin{center}
\vspace{-0.0cm}
\includegraphics[trim=0.0cm 0.0cm 0.0cm 0.0cm ,clip, angle=0, scale=0.6]{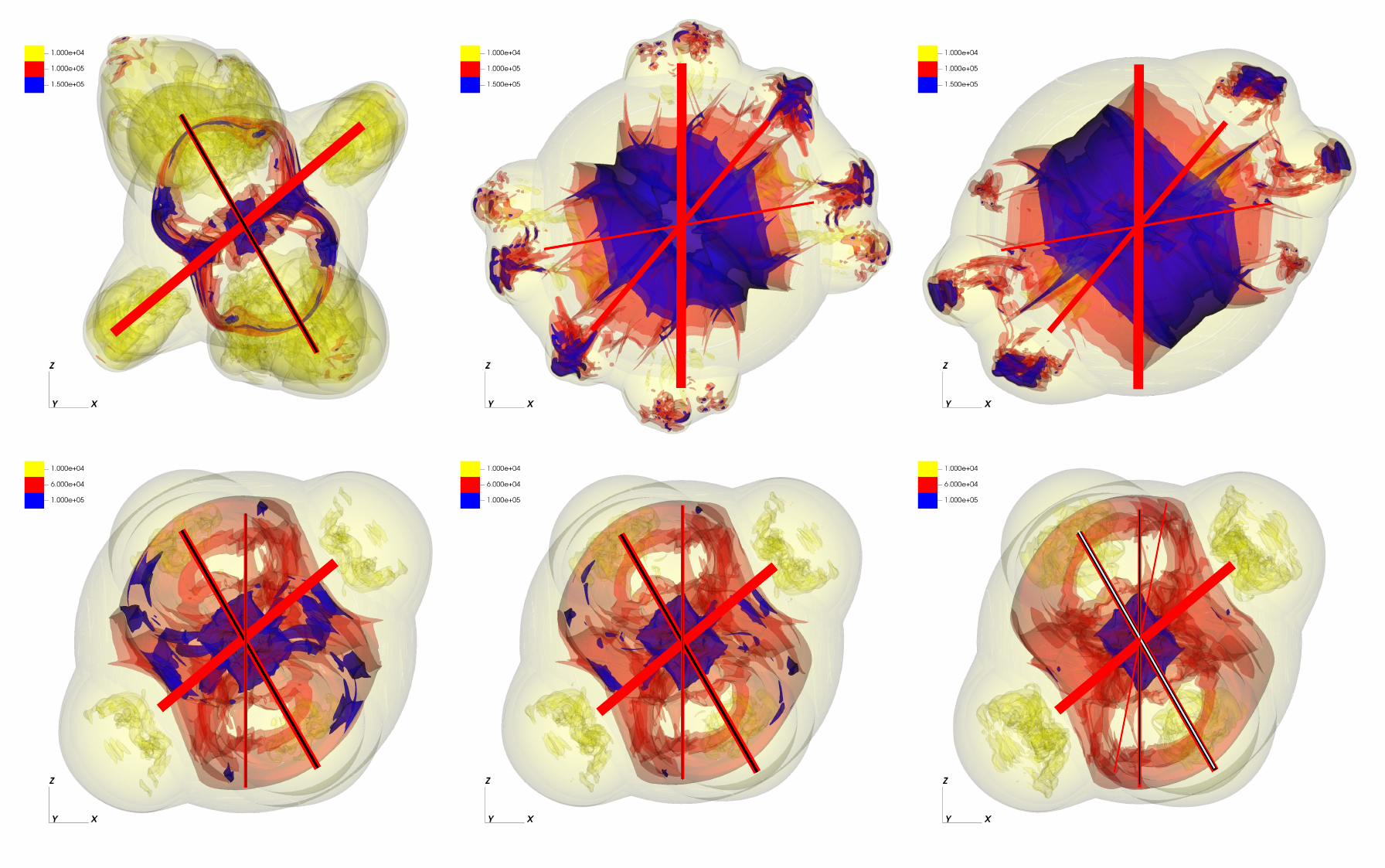} 
\vspace{-0.0cm}
\caption{Similar to Figure \ref{Fig:contouryx} but as seen from the $y$ direction. On each panel, we added the projection of the jets' axes of the simulation on this plane using straight lines. Thicker lines are earlier jets and thinner lines are late jets. In cases where two jets' axes are projected on one another, the thinner (later) one is in a different color. On the lower right panel, three jets are projected at each other. Note that dense clumps might form along the jets' axes or to their sides. 
}
\label{Fig:contourzx}
\end{center}
\end{figure*}

The 3D density surface maps, Figures \ref{Fig:contouryx} - \ref{Fig:contourzx}, clearly present the point-symmetric morphologies by emphasizing pairs of structural features, particularly bubbles and clumps. 
Bubbles are prominent, as example, in the lower right panel of Figure \ref{Fig:contouryx} for simulation B7 (two opposite white areas), and in the three lower panels of Figure \ref{Fig:contourzx}, simulations A5, B5, and B7 (the pairs of yellow areas surrounded by red rims). The upper left panels of Figures \ref{Fig:contouryx} - \ref{Fig:contourzx} show the large bubbles of simulation A3 that we present in Figure \ref{Fig:densA3}. 

Pairs of clumps are prominent in simulations P1 and P2 that we present in the upper middle and upper right panels of Figures \ref{Fig:contouryx} - \ref{Fig:contourzx}, particularly in Figure \ref{Fig:contourzx}. Inner pairs of clumps are also prominent, as small blue areas in the larger red zones, in simulations A5 and B5, as seen in the lower left and middle panels in Figures \ref{Fig:contouryx} - \ref{Fig:contourzx}.
The upper middle panel of Figure \ref{Fig:contourzx} is for simulation P1 with three pairs of jets. However, we see five pairs of clumps, and each of the 10 clumps is fragmented to smaller knots, due to instabilities that we discuss in Section \ref{sec:RTI}. The finding that three pairs of jets can form five pairs of clumps is highly significant when analyzing observations. \citet{SokerShishkin2025Vela} who identified seven pairs of clumps in the Vela CCSN remnant, already pointed out that two pairs result from one pair of jets. \citet{BearSoker2025CasA} who identified the point-symmetric structure of Cassiopeia A with nine pairs of structural features, proposed that a pair of jets can form two or more pairs of clumps and filaments. Our results solidify the claim from observations that a pair of jets can form two or more pairs of clumps and filaments. Or, more accurately, the interaction between two pairs of jets can form more than two pairs of clumps.  
\subsection{Mimicking observations}
\label{subsec:Observtions}

Observations show the point-symmetric morphologies of CCSN remnants long after the shock breaks from the star, and the ejecta is optically thin. In our simulations, the jet-driven outflow is still inside the star. Nonetheless, we mimic the observed morphology by integrating the density square along the line of sight. This approach is commonly used in astrophysical simulations to generate relative emission maps, as the emissivity of optically thin nebulae is often proportional to the square of the density. The numerical emission measure is 
\begin{equation}
    {\rm EM}(U,W) = \int \rho^2(x,y,z) \, dL,
    \label{eq:proj}
\end{equation}
where $\rho(x,y,z)$ is the density, $(U,W)$ are the coordinates on the plane of the sky perpendicular to the line of sight, and the integration is along the line of sight $L$. 

In Figure \ref{Fig:projP1} we present the numerical emission measure map of simulation P1 along the $y$ axis, the same direction of the upper middle panel in Figure \ref{Fig:contourzx}. The numerical emission measure map, which best mimics observed maps, clearly presents the formation of five pairs of clumps, each composed of several knots (smaller clumps). The formation of five pairs of clumps by three pairs of jets that explode the core, is one of the major results of our study.  
\begin{figure}[h]
\begin{center}
\includegraphics[trim=2.0cm 6.2cm 2.0cm 6.25cm ,clip, angle=270, scale=0.50]{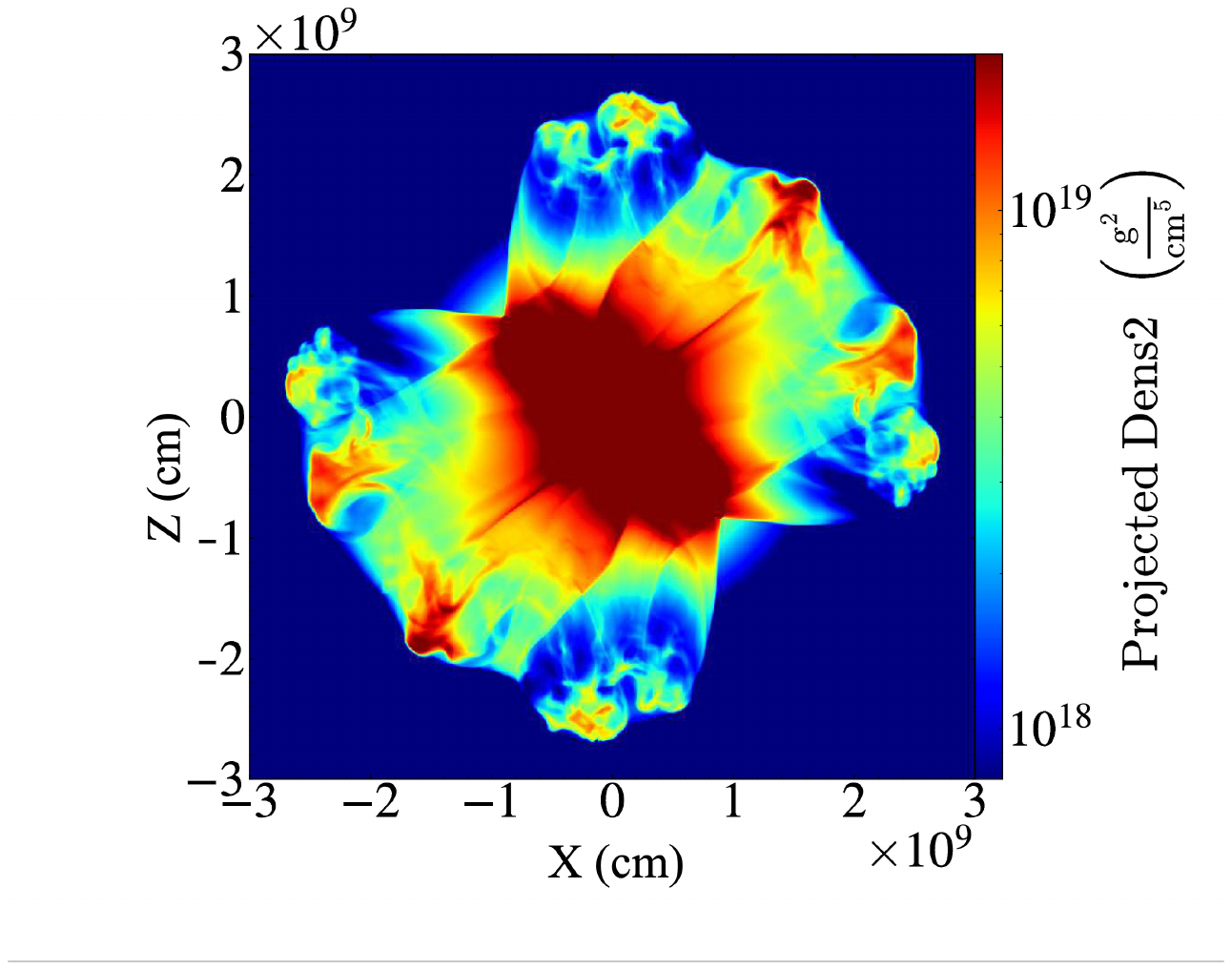} 
\vspace{-0.0cm}
\caption{The numerical emission measure map ${\rm EM}(X,Z)$ for simulation P1, according to equation (\ref{eq:proj}), where the integration is along the $y$ direction, i.e., $dL=dy$. EM maps best mimic observed morphologies. As the 3D map of simulation P1 in the upper middle panel in Figure \ref{Fig:contourzx} shows, the simulation with three pairs of jets forms five pairs of clumps, and each clump is fragmented to several knots (smaller clumps). 
Axes are from $-30,000 \km$ to $30,000 \km$, and colors span the range of ${\rm EM}(X,Z) =8 \times 10^{17} \g^2 \cm^{-5}$ (deep blue) to  ${\rm EM}(X,Z) =2 \times 10^{19} \g^2 \cm^{-5}$ (deep red). 
}
\label{Fig:projP1}
\end{center}
\end{figure}

Figure \ref{Fig:projA3} and \ref{Fig:projA5} present the numerical emission measure maps of simulation A3 and A5 that we also present in Figures \ref{Fig:densA3} and \ref{Fig:densA5}, respectively. These maps are the closest among the figures we present here to real observations. 
\begin{figure}
\begin{center}
\vspace{-0.0cm}
\includegraphics[trim=7.0cm 6.2cm 7.34cm 5.2cm ,clip, scale=1.15]{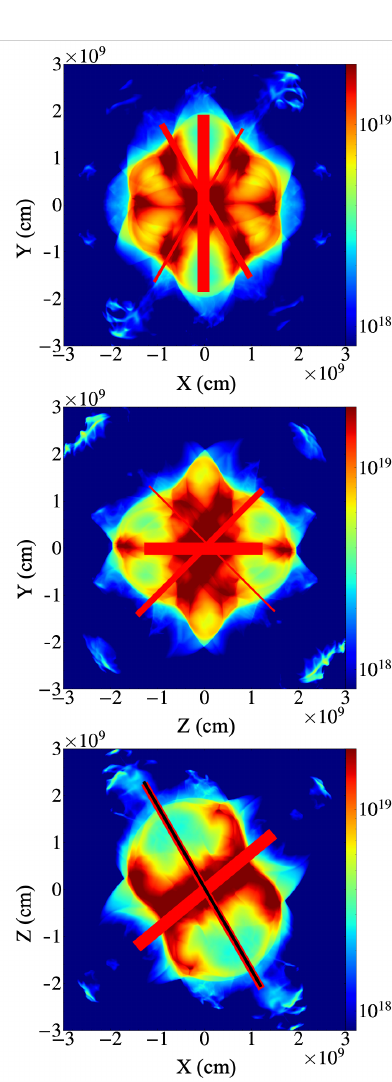} 
\vspace{-0.0cm}
\caption{Similar to Figure \ref{Fig:projP1} but of simulation A3 and along three lines of sight, from top to bottom, along the $z$, $x$, and $y$ directions. The straight lines are the projection of the jets' axes, same as in Figure \ref{Fig:contourzx}. 
}
\label{Fig:projA3}
\end{center}
\end{figure}
\begin{figure}
\begin{center}
\vspace{-0.0cm}
\includegraphics[trim=7.0cm 6.1cm 7.34cm 5.2cm ,clip, scale=1.15]{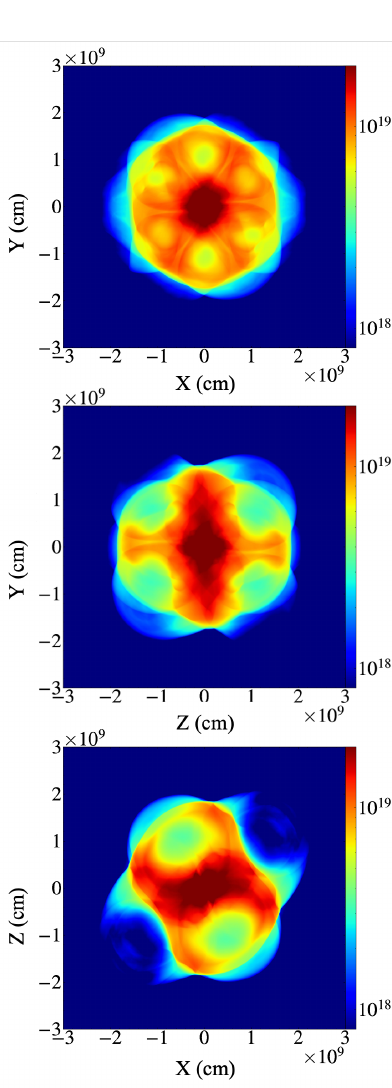} 
\vspace{-0.0cm}
\caption{
Similar to Figure \ref{Fig:projA3} but for simulation A5. 
}
\label{Fig:projA5}
\end{center}
\end{figure}

The emission measure maps in Figures \ref{Fig:projP1} - \ref{Fig:projA5} (and the maps of the other simulations we do not present here) reveal the following observational properties that result from the JJEM.
\begin{enumerate}
    \item Pairs of jets form a clear and robust point-symmetric morphology. 
    \item A simulation might reveal different morphologies from different directions. 
    \item Clumps can form between jets-inflated bubbles (or lobes) and at the tip of the jets. 
    \item On average, a pair of jets might form two or more clumps or filaments. 
    \item Some clumps break into smaller clumps that we term knots. This breaking results from the instabilities we study in Section \ref{sec:RTI}.
    \item Some maps show the formation of several pairs of clumps. The three pairs of large clumps in the upper panel of Figure \ref{Fig:projA3}, qualitatively resemble the point-symmetric structure in the Doppler map of SNR 0540-69.3 that \citet{Soker2022SNR0540} identified as four pairs of clumps (for the observation see \citealt{Larsonetal2021}).   \item The middle panel of Figure \ref{Fig:projA3} shows an elongated structure with two opposite clumps at the tips. Many jet-shaped objects have such structures in clusters of galaxies, planetary nebulae, and the SNR W49B \citep{SokerShishkin2025W49}. 
    \item Usually, bubbles are inflated by jets. However, in some cases, small bubbles might appear between jet axes. These are the two opposite observed bubbles along  $Z=0$ in the middle panel of Figure \ref{Fig:projA3}. 
    \item An `H-shaped' structure dominates the morphology in the lower panel of Figure \ref{Fig:projA3} (in deep red), with two opposite outer faint clumps along the symmetry axis (the axis parallel to the legs of the H and through the center). This strengthens the claim that jets shaped the H-shaped structure of SNR W49B with jets parallel to the legs of the `H' (e.g., \citealt{BearSoker2017PNSNR, Akashietal2018, Siegeletal2020, GrichenerSoker2023, SokerShishkin2025W49}).
    \item The upper panel of Figure \ref{Fig:projA5} shows a morphology of three pairs of bubbles. These are jet-inflated bubbles. The radioactive decay of nickel can also form bubbles (nickel bubbles, e.g., \cite{Orlandoetal2025}). The decay of nickel and the inflation of bubbles occur long after the explosion. Nickel bubbles will form as two opposite bubbles if a pair of jets ejects two opposite clumps of nickel.    
\end{enumerate}

The six simulations demonstrate the rich variety of point-symmetric morphologies that jets can shape, some of which resemble observed CCSN remnants, although we were not aiming to replicate any specific observation. 

\section{Results: Rayleigh-Taylor instabilities and vortices}
\label{sec:RTI}

As a jet deposits energy into the jet-core interaction region, it forms a high-pressure volume that expands. As a result of this expansion, the density within the bubble decreases below that of its surroundings, while its pressure remains higher. Zones of opposite pressure and density gradients are prone to the Rayleigh-Taylor instability (RTI).    
To locate these zones and their growth rate, we refer to the quantity 
\begin{equation}
f_{st} \equiv \frac{1}{\rho} \sqrt{\lvert \vec{\nabla}P \cdot \vec{\nabla}\rho \rvert} \ \text{sgn}(\vec{\nabla}P\cdot\vec{\nabla}\rho), 
\label{eq:rt}      
\end{equation}
Where $\rho$ is the density, $P$ is the pressure, and ${\rm sgn} (\vec{\nabla}P\cdot\vec{\nabla}\rho)$ is the sign of the product of pressure and density gradients. 
If $f_{st}$ is negative, the region is unstable with a typical growth rate of $-f_{st}$, and a typical growth time of $-1/f_{st}$.

The upper row panels of Figure \ref{Fig:rtA3} present maps of $f_{\rm st}$ in units of $\s^{-1}$ for simulation A3 at $t=1.36 \s$ in three planes of the grid. In the deep red zones $f_{\rm st}>0$ and the red areas are RTI stable. The yellow to deep blue zones are RTI unstable, with the fastest growing rate in the deep blue zones, a growth time of about 1 second. Although we present the results at the final time of the simulation, the RTI persists throughout the entire jet-core interaction. The growth time is shorter than the simulation time, allowing the RTI to transition into the non-linear regime. The RTI explains the breakup of clumps to knots (smaller clumps; Section \ref{sec:Density}). 
The unstable zones are thin surfaces that appear as thin filaments on the planes. The unstable surfaces occupy a small fraction of the volume. As with water lying on oil, unstable zones are located near the interface between the two media and may have a small filling factor, but they still contribute to the development of large-amplitude instabilities. Therefore, even if the RTI unstable zones with short time growth have a small filling factor, the RTI influences the small-scale structure of the ejecta, e.g., breaking clumps and filaments, because the RTI growth time is shorter than the simulation time. The panels in the upper row of Figure \ref{Fig:rtA5} present the RTI maps of simulation A5 at $t=1.94 \s$. 
\begin{figure*}[t]
\begin{center}
\vspace{-0.0cm}
\includegraphics[trim=0.0cm 8.0cm 0.0cm 8.0cm ,clip, scale=0.85]{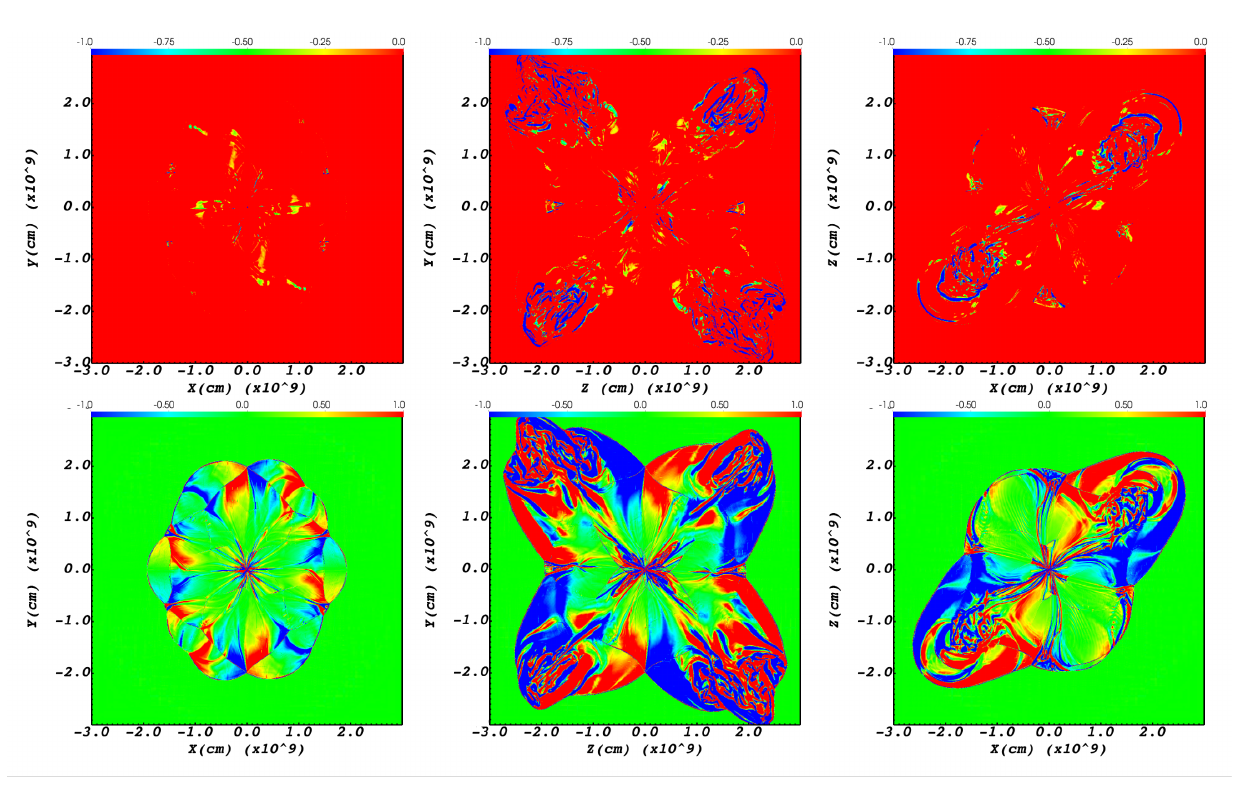} 
\vspace{-0.0cm}
\caption{RTI growth rate and vorticity color maps of simulation A3 at $t=1.36\s$ in the planes, from left to right, $z=0$, $x=0$, and $y=0$. The upper panels show $f_{st}$, calculated using equation (\ref{eq:rt}). Deep-red areas are RTI stable, while others are unstable with a typical growth time of $-1/f_{st}$. Color bar is from $-1 \s^{-1}$ (deep blue) to $\ge 0$ (deep red). The lower panels show the vorticity component perpendicular to the given plane, $(\nabla\times\vec{v})_\perp$. The color bar runs from $-1 \s^{-1}$ (deep blue clockwise) to $1 \s^{-1}$ (deep red; counterclockwise). 
}
\label{Fig:rtA3}
\end{center}
\end{figure*}
\begin{figure*}[t]
\begin{center}
\vspace{-0.0cm}
\includegraphics[trim=0.0cm 8.0cm 0.0cm 8.0cm ,clip, scale=0.85]{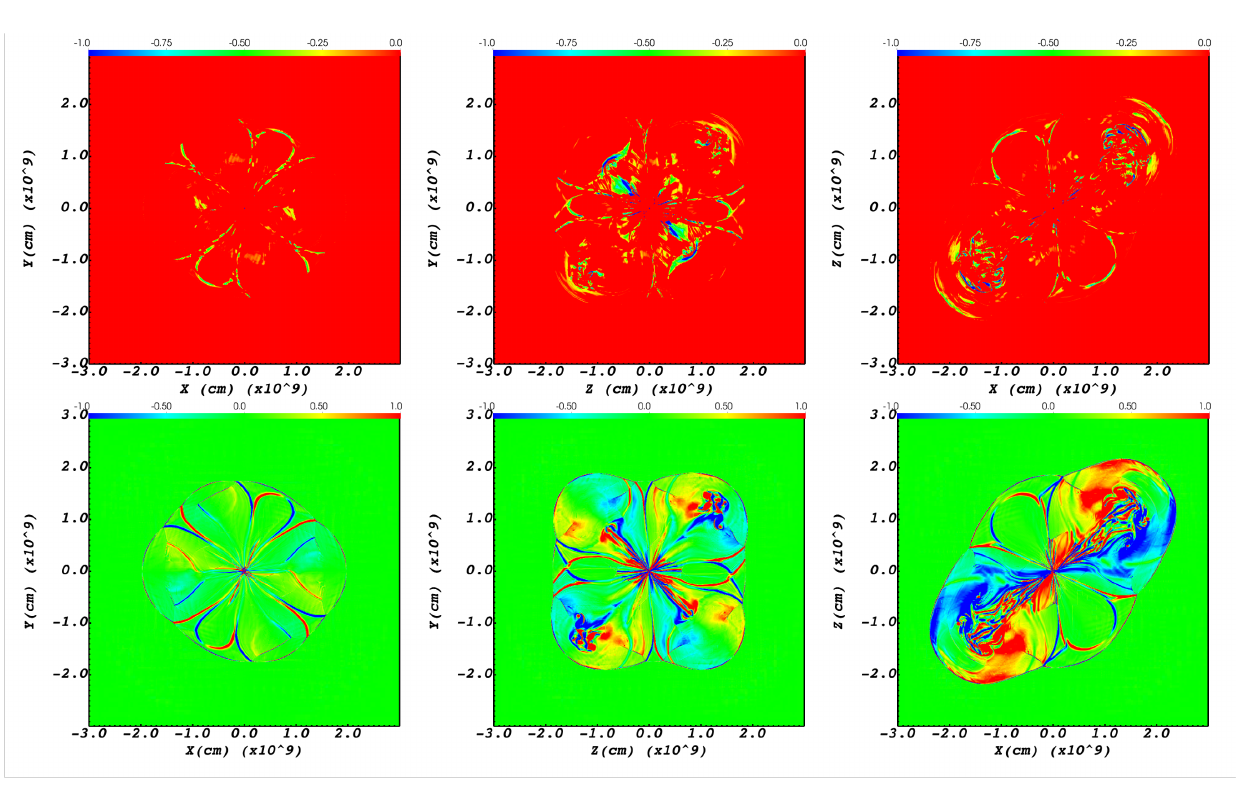} 
\vspace{-0.0cm}
\caption{
Similar to Figure \ref{Fig:rtA3} but for simulation A5 and at $t=1.94 \s$.
}
\label{Fig:rtA5}
\end{center}
\end{figure*}

In the RTI, low-density fingers protrude into the high-density low-pressure zone, and high-density fingers protrude into the low-density high-pressure zone. The result is the formation of vortices on small scales in RTI unstable zones. On the other hand, the interaction of the jets with the core material forms mainly vortices on large scales, i.e., of the order of the scales of the bubbles the jets inflate. This difference is best seen in the middle panels of Figure \ref{Fig:rtA3}.
The RTI unstable zones that are zones with blue stripes on the upper-middle panel of Figure \ref{Fig:rtA3}, have red-blue stripes in the middle-lower panel that show the vorticity component perpendicular to the plane, i.e., $(\nabla\times\vec{v})_\perp$. These are small-scale vortices. The continuous large blue and red areas in the middle-lower panel are large-scale vortices induced by the backflow that the jets form; comparing the two middle panels, shows these areas to be RTI stable. 
The role of instabilities and vortices in nucleosynthesis during the explosion process and the mixing of elements are topics for a future study in this series. The lower rows of Figures \ref{Fig:rtA3} and \ref{Fig:rtA5} show the vortex values in the other planes of simulation A3 and A5. 

In Figure \ref{Fig:v_all} we present velocity maps in two planes of simulation A3 and A5 to further demonstrate the large-scale vortices that the jets induce. The colors represent the magnitude of the velocity (red indicating high velocity), and the arrows are aligned with the flow direction. In the lobes, the highest outward velocity is along the axis of symmetry. The expansion velocity is lower on the sides of the symmetry axis of a lobe/bubble. In the frame of the expanding lobes, there are large vortices with a flow from the center out and then back, i.e., a backflow.  
\begin{figure*}[t]
\begin{center}
\vspace{-0.5cm}
\includegraphics[trim=0.0cm 0.0cm 0.0cm 0.0cm ,clip, scale=0.9]{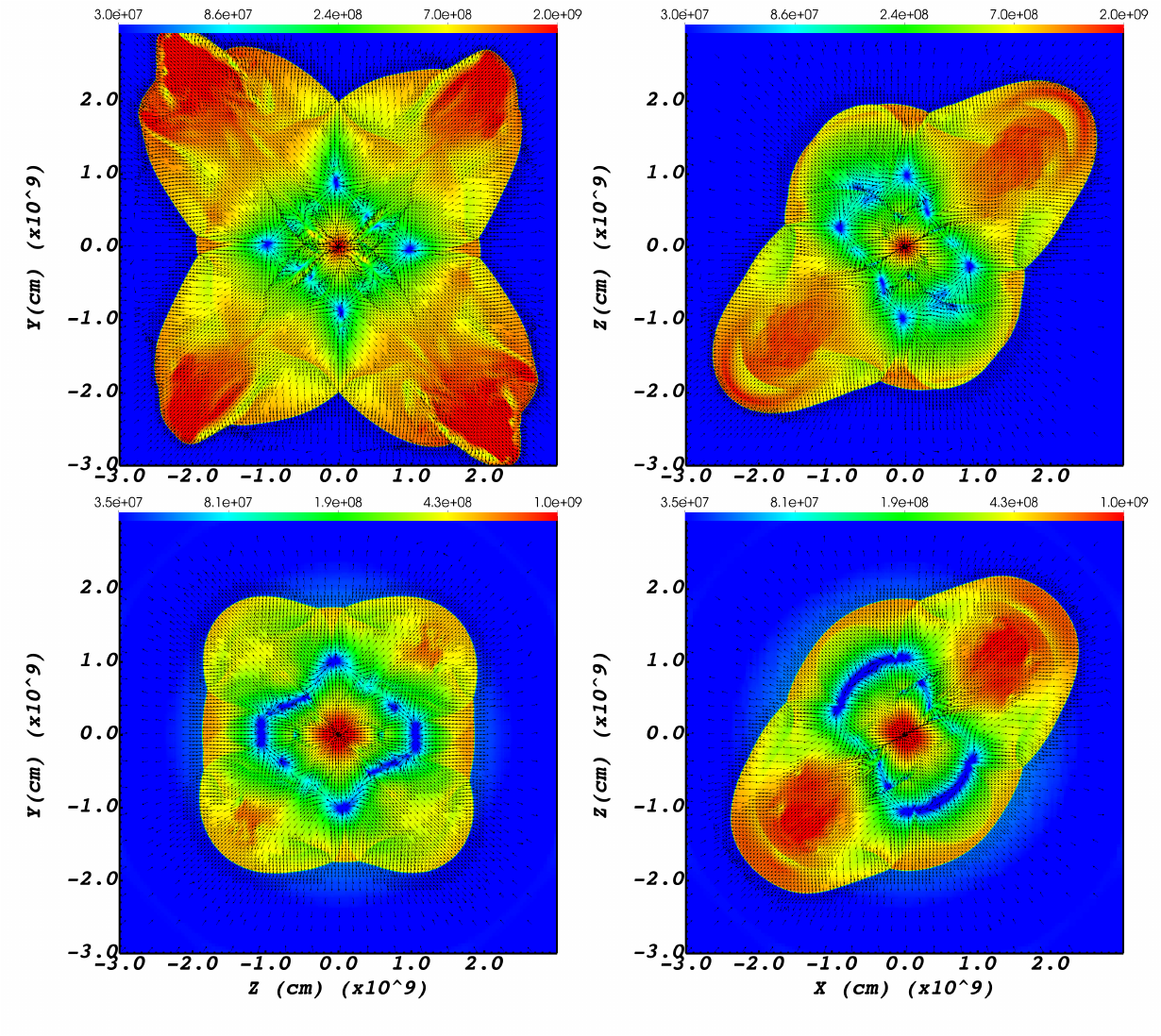} 
\vspace{-0.9cm}
\caption{
Velocity magnitude and direction color maps of simulation A3 (upper panels) at $t=1.36\s$ and simulation A5 (lower panels) at $t=1.94\s$, and in the $x=0$ plane (left) and $y=0$ plane (right). The color bar indicates the magnitude of the velocity; In the upper panels, the color bar is from $3\times10^7 \cm \s^{-1}$ (deep blue) to $2\times10^9 \cm \s^{-1}$ (deep red), and in the lower panels from $3.5\times10^7 \cm \s^{-1}$ to $10^9 \cm \s^{-1}$. The black arrows indicate the direction of the velocity at that point. Note that the backflow near the center leads to further accretion and more jets, which we will study in a future paper.  
}
\label{Fig:v_all}
\end{center}
\end{figure*}

The velocity maps of Figure \ref{Fig:v_all} show that near the center, the material flows inward. We do not treat the center accurately in this study. Firstly, the newly born NS has a wind that can push material out. Secondly, and more importantly, the backflow near the center may lead to further accretion, which in turn launches more jets. In a future study, we will include later jets and the expansion of the forward shock out from the core.  

The velocity maps of simulation A3 in the upper panels of Figure \ref{Fig:v_all} show pairs of expanding lobes. The shocked core's material occupies the outer volume of these lobes, while the original jets' material is inside these lobes. To demonstrate this, we present in Figure \ref{Fig:tracers} the tracer of the jets of simulation A3 in the same planes as the velocity maps. 
The tracer is a numerical quantity that traces the flow of a designated volume. In the present case, it is the material of the jets. At jets' launching, the tracer in the jet equals 1, i.e., pure jets' material. Over time, the value of the tracer in a cell indicates the fraction of mass that originated from the jets; the tracer can assume any value between 0 and 1. The size of the lobes in the tracer maps of Figure \ref{Fig:tracers} is much smaller than the expanding high-velocity lobes of the upper panels of Figure \ref{Fig:v_all}. This indicates that the original jets' material is located deep within the rapidly expanding lobes. 
\begin{figure}[h]
\begin{center}
\vspace{-0.0cm}
\includegraphics[trim=5.0cm 6.0cm 5.0cm 5.8cm ,clip, scale=0.8]{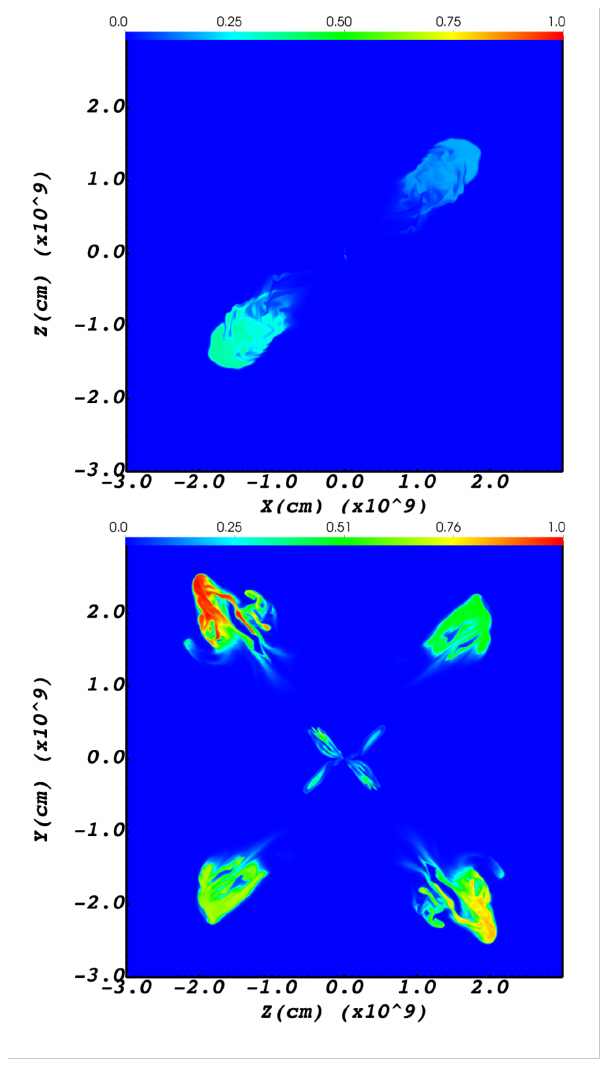} 
\vspace{-0.9cm}
\caption{Jets' tracer color map of simulation A3 at $t=1.36\s$ in the $y=0$ plane (upper panel) and $x=0$ plane (lower panel). The tracer is the fraction of material originating in the jets in the given cell. The color bar is from 0 (deep blue), which indicates that no material comes from the jets, to 1 (deep red), which indicates that in that cell all material originated in the jets. 
}
\label{Fig:tracers}
\end{center}
\end{figure}

\section{Summary}
\label{sec:Summary}

To explain about a dozen point-symmetric CCSN remnants, we conducted six 3D hydrodynamical simulations of the JJEM. In these simulations, we launch several pairs of jittering jets in varying directions inside a collapsing massive stellar core model. We examined the morphologies that the jittering jets shape. In this study, we do not follow the backflow or the NS wind, so we did not analyze the grid's inner $\simeq 15 \%$. 

The density structures in Figures \ref{Fig:densA3} - \ref{Fig:contourzx} show many point-symmetric morphologies, depending on the simulation parameters and the viewing angle. To mimic observations during the CCSN remnant phase, when the ejecta is optically thin, we present the numerical emission measure (equation \ref{eq:proj}) of three simulations in Figure \ref{Fig:projP1} - \ref{Fig:projA5}. The density and numerical emission measure maps show that jets form pairs of bubbles, ears, clumps, and filaments and a rich variety of point-symmetric morphologies capable of accounting for the observed morphologies of most point-symmetric CCSN remnants with pairs of lobes (e.g., G321.3–3.9; SN 1987A), and pairs of clumps and filaments (e.g, SNR 0540-69.3; Vela; SN 1987A; 
Cassiopeia A; the Crab Nebula; W49B; for references see Section \ref{sec:Introduction}). 

We summarized the results of our simulations in a list of points in Section \ref{subsec:Observtions}. We consider most significant our finding that on average a pair of jets might form more than one pair of clumps, that small bubbles might form not only along the jets' axis, that RTI might break clumps into smaller clumps (knots), and that jets can form an `H-shaped' CCSN remnant like W49B. These are significant results because they explain specific observed properties of some CCSN remnants. Our study further supports the JJEM as a viable mechanism of CCSNe.

In future studies we will follow the explosion to break out from the star, the launching of late jets due to the backflow (see Figure \ref{Fig:v_all} for the flow structure), the nucleosynthesis as the jets interact with the core material, and the role of vortices and RTIs (Figures \ref{Fig:rtA3} and \ref{Fig:rtA5}) in distributing and mixing the newly synthesis elements. Our results will lead the search for more jet signatures in CCSNe and CCSN remnants. 
For example, we expect that when a jet breaks out from the star, possibly late jets, it will experience a shock and, for the lack of dense ambient gas, will lose collimation. This is observed in jets in galaxies and planetary nebulae. In CCSNRs, the outer uncollimated jet's material is referred to as a blowout, and was studied in the frame of the JJEM for the Cygnus Loop and SNR G309.2–00.6 by \cite{ShishkinKayeSoker2024}, and for SNR G0.9+0.1 by \cite{Soker2025G09}.

Words are in place here on the general approach of our study, which 
is very common in studying active galactic nuclei (AGN) feedback and in shaping planetary nebulae. It is based on these ingredients. 
(1) Strong observational indications of jet activity at present or in the past, like presently active jets, outcome of past jets (like pairs of bubbles and lobes), or gamma rays in gamma ray bursts.
We have the point-symmetric morphologies of more than ten CCSN remnants (Section \ref{sec:Introduction}), and more CCSN remnants with one pair of ears, which indicate jet activity. 
(2) The presence of an accreting compact object, like a super-massive black hole in an AGN, a companion to the progenitor of a planetary nebula, or a stellar mass black hole in gamma ray bursts.
CCSNe form an NS or a black hole at the center that accretes more core material.
(3) Manual jet launching without resolving the compact object's vicinity. Namely, the jets are launched from a distance much larger than the accretion disk that launches them. The numerical launching of jets does not refer to the physical mechanism by which the accretion disk launches the jets, which is a difficult and unsolved process. Particularly relevant to us, a recent study finds that current numerical simulations lack the resolution to accurately capture magnetic field reconnection events, which are crucial for jet launching in CCSNe \citep{Soker2025Learning}. 

Identifying point-symmetric morphologies in the images of CCSN remnants in the last two years (Section \ref{sec:Introduction}) has been a breakthrough in our understanding of the explosion mechanism of CCSNe.  By reproducing the general point-symmetric structures observed in these CCSN remnants, our study encourages further simulations of the JJEM aiming at reproducing more morphologies and other CCSN properties, e.g., distributions of radioactive isotopes.

\section*{Acknowledgements}
We thank Dmitry Shishkin for helpful comments and discussions, and an anonymous referee for helpful comments. 
Amir Michaelis is supported by the European Union’s Horizon 2020 research and innovation program under grant agreement No 865932-ERC-SNeX.
We acknowledge the Ariel HPC Center at Ariel University for providing computing resources that have contributed to the research results reported in this paper.



%




\appendix
\section{Initial parameters of simulations}
\label{subsec:initial_param}

Tables \ref{Tab:Table3} - \ref{Tab:Table6} present the parameters of the four simulations that we did not present in the main text. 
\begin{table}[h]
\begin{center}
  \caption{Initial parameters of simulation P1}
    \begin{tabular}{| p{2.0cm} | p{1.1cm} | p{1.1cm}| p{1.1cm} |}

\hline
Jets' pair & $1$ & $2$ & $3$ \\
\hline
$t_{\rm j}({\rm start})\ [\rm s]$ & $0.00$ & $0.05$ & $0.1$\\
\hline
$t_{\rm j} {\rm(end)}\ [\rm s]$ & $0.05$ & $0.1$ & $0.25$\\
\hline
$\theta_{\rm j}\ [\rm deg]$ & $0$ & $40$ & $80$ \\
\hline
$\phi_{\rm j}\ [\rm deg]$ & $0$ & $0$ & $0$ \\
\hline
$\alpha_{\rm j}\ [\rm deg]$ & $10$ & $10$ & $10$ \\
\hline
$\dot{m}_{\rm j}\ [10^{32} \g \s^{-1}]$ & $4$ & $4$ & $4$ \\
\hline
$E_{\rm k,2j}\ [10^{50} \erg]$ & $5$ & $5$ & $15$ \\
\hline

\end{tabular}
  \\
\label{Tab:Table3}
\end{center}
\begin{flushleft}
\small 
Notes: Similar to Table \ref{Tab:Table1} but for simulation P1, composed of three pairs of jets launched sequentially in the $y=0$ plane. Note that the last pair lasts longer than the other pairs. The total kinetic energy carried by jets in this simulation is $E_{\rm k,tot} ({\rm P1}) = 2.5 \times 10^{51}\ \rm erg$.   
\end{flushleft}
\end{table}
\begin{table}[h]
\begin{center}
  \caption{Initial parameters of simulation P2}
    \begin{tabular}{| p{2.0cm} | p{1.1cm} | p{1.1cm}| p{1.1cm} |}

\hline
Jets' pair& $1$ & $2$ & $3$ \\
\hline
$t_{\rm j}({\rm start})\ [\rm s]$ & $0.00$ & $0.05$ & $0.1$\\
\hline
$t_{\rm j} {\rm(end)}\ [\rm s]$ & $0.05$ & $0.1$ & $0.25$\\
\hline
$\theta_{\rm j}\ [\rm deg]$ & $0$ & $40$ & $80$ \\
\hline
$\phi_{\rm j}\ [\rm deg]$ & $0$ & $0$ & $0$ \\
\hline
$\alpha_{\rm j}\ [\rm deg]$ & $40$ & $10$ & $10$ \\
\hline
$\dot{m}_{\rm j}\ [10^{32} \g \s^{-1}]$ & $4$ & $4$ & $4$ \\
\hline
$E_{\rm k,2j}\ [10^{50} \erg]$ & $5$ & $5$ & $15$ \\
\hline

\end{tabular}
  \\
\label{Tab:Table4}
\end{center}
\begin{flushleft}
\small 
Notes: Similar to Table \ref{Tab:Table1} but for simulation P2. P2 is identical to P1 except for the larger half-opening angle $\alpha_{\rm j}$ of the first pair of jets. The total kinetic energy carried by jets in this simulation is $E_{\rm k,tot} ({\rm P2}) = 2.5 \times 10^{51}\ \rm erg$.   
\end{flushleft}
\end{table}
\begin{table}[h]
\begin{center}
  \caption{Initial parameters of simulation B5}
    \begin{tabular}{| p{2.0cm} | p{0.7cm} | p{0.7cm}| p{0.7cm}| p{0.7cm} | p{0.7cm} |}

\hline
Jets' pair& $1$ & $2$ & $3$ & $4$ & $5$ \\
\hline
$t_{\rm j}({\rm start})\ [\rm s]$ & $0.00$ & $0.07$ & $0.14$ & $0.21$ & $0.28$ \\
\hline
$t_{\rm j} {\rm(end)}\ [\rm s]$ & $0.05$ & $0.12$ & $0.19$ & $0.26$ & $0.413$ \\
\hline
$\theta_{\rm j}\ [\rm deg]$ & $50$ & $50$ & $50$ & $0$ & $0$ \\
\hline
$\phi_{\rm j}\ [\rm deg]$ & $0$ & $120$ & $240$ & $0$ & $20$\\
\hline
$\alpha_{\rm j}\ [\rm deg]$ & $10$ & $10$ & $10$ & $10$ & $5$ \\
\hline
$\dot{m}_{\rm j}\ [10^{32} \g \s^{-1}]$ & $1.8$ & $1.8$ & $1.8$ & $1.8$ & $1.8$ \\
\hline
$E_{\rm k,2j}\ [10^{50} \ \rm erg]$ & $2.25$ & $2.25$ & $2.25$ & $2.25$ & $6$ \\
\hline

\end{tabular}
  \\
\label{Tab:Table5}
\end{center}
\begin{flushleft}
\small 
Notes: Similar to Table \ref{Tab:Table1} but for simulation B5. B5 is identical to A5 except for the smaller half-opening angle $\alpha_{\rm j}$ of the last pair of jets. The total kinetic energy carried by the jets in this simulation is $E_{\rm k,tot} ({\rm B5}) = 1.5 \times 10^{51}\ \rm erg$.     
\end{flushleft}
\end{table}
\begin{table}[h]
\begin{center}
  \caption{Initial parameters of simulation B7}
    \begin{tabular}{| p{2.0cm} | p{0.6cm} | p{0.6cm}| p{0.6cm}| p{0.6cm} | p{0.6cm} |p{0.6cm}| p{0.6cm}|}

\hline
Jets' pair& $1$ & $2$ & $3$ & $4$ & $5$ & $6$ & $7$ \\
\hline
$t_{\rm j}({\rm start}) [\rm s]$ & $0.00$ & $0.07$ & $0.14$ & $0.21$ & $0.28$ & $0.35$ & $0.42$ \\
\hline
$t_{\rm j} {\rm(end)} [\rm s]$ & $0.05$ & $0.12$ & $0.19$ & $0.26$ & $0.33$ & $0.40$ & $0.47$\\
\hline
$\theta_{\rm j} [\rm deg]$ & $50$ & $50$ & $50$ & $0$ & $0$ & $50$ & $30$ \\
\hline
$\phi_{\rm j} [\rm deg]$ & $0$ & $120$ & $240$ & $0$ & $20$ & $240$ & $70$ \\
\hline
$\alpha_{\rm j} [\rm deg]$ & $10$ & $10$ & $10$ & $10$ & $20$ & $10$ & $10$ \\
\hline
$\dot{m}_{\rm j} [10^{32} \g \s^{-1}]$ & $1.8$ & $1.8$ & $1.8$ & $1.8$ & $1.8$ & $1.8$ & $1.2$ \\
\hline
$E_{\rm k,2j} [10^{50} \ \rm erg]$ & $2.25$ & $2.25$ & $2.25$ & $2.25$ & $2.25$ & $2.25$ & $1.5$ \\
\hline

\end{tabular}
  \\
\label{Tab:Table6}
\end{center}
\begin{flushleft}
\small 
Notes: Similar to Table \ref{Tab:Table1} but for simulation B7, composed of seven pairs of jets. The first five pairs are identical to those in simulation A5. All pairs have the same active time of jet-launching. Note that the last pair has a lower mass loss rate $\dot{m_{\rm j}}$ than the other pairs. The total kinetic energy carried by jets in this simulation is $E_{\rm k,tot} ({\rm B7}) = 1.5 \times 10^{51}\ \rm erg$.       
\end{flushleft}
\end{table}

\end{document}